\documentclass[]{article}
\usepackage{graphicx}
\usepackage{color}
\usepackage{listings}
\usepackage{fullpage}
\usepackage{amsmath}
\usepackage[utf8x]{inputenc}
\usepackage{import}
\usepackage{setspace}
\usepackage{authblk}
\usepackage{hyperref}
\definecolor{lightgray}{gray}{0.5}
\usepackage{setspace}
\usepackage{color}

\begin{document}

\title{Evolution of semantic networks in biomedical texts}
\author[1,2]{Lucy R. Chai}
\author[1,3,4,5,6]{Danielle S. Bassett}
\affil[1]{Department of Bioengineering, School of Engineering and Applied Sciences, University of Pennsylvania, Philadelphia, PA 19104 USA}
\affil[2]{Program in Machine Learning, Speech and Language Technology, Department of Engineering, University of Cambridge, Cambridge, UK CB2 1ST}
\affil[3]{Department of Physics \& Astronomy, College of Arts and Sciences, University of Pennsylvania, Philadelphia, PA 19104 USA}
\affil[4]{Department of Electrical \& Systems Engineering, School of Engineering and Applied Sciences, University of Pennsylvania, Philadelphia, PA 19104 USA}
\affil[5]{Department of Neurology, Perelman School of Medicine, University of Pennsylvania, Philadelphia, PA 19104 USA}
\affil[6]{To whom correspondence should be addressed: dsb@seas.upenn.edu}

\maketitle

\begin{abstract}

Language is hierarchically organized: words are built into phrases, sentences, and paragraphs to represent complex ideas. A similar hierarchical structure is observed across many other biological, electronic, and transportation networks supporting complex functions. Here we ask whether the organization of language in written text displays fractal hierarchical architecture. Specifically, we test two hypotheses: (i) that the structure of the exposition in scientific research articles displays the Rentian scaling principle, which marks hierarchical fractal-like structure, and (ii) that the exponent of the scaling law changes as the article is revised to maximize information transmission. Using 32 scientific manuscripts -- each containing between three and 26 iterations of revision -- we construct semantic networks in which nodes represented unique words in each manuscript, and in which edges connect nodes if two words appeared within the same 5-word window. We show that these semantic networks modeling the content of scientific articles display clear Rentian scaling, and that the Rent exponent varies over the publication life cycle, from the first draft to the final revision. Furthermore, we observe that manuscripts fell into three clusters in terms of how the scaling exponents changed across drafts: exponents rising over time, falling over time, and remaining relatively stable over time. This change in exponent reflects the evolution in semantic network structure over the manuscript revision process, highlighting a balance between network complexity, which increases the exponent, and network efficiency, which decreases the exponent. Lastly, the final value of the Rent exponent is negatively correlated with the number of authors. Taken together, our results suggest that semantic networks reflecting the structure of exposition in scientific research articles display striking hierarchical architecture that arbitrates trade-offs between competing constraints on network organization, and that this arbitration is navigated differently depending on the social environment characteristic of the collaboration. 

\end{abstract}

\newpage

\section{Introduction}
\doublespacing
Many information transmission networks -- from artificial or natural neural networks to email communication networks and the internet -- share a number of core organizational properties. Such properties include community structure (or more colloquially \emph{modularity}) and relatively small shortest path lengths (or hop distances) \cite{lancichinetti2010characterizing,sporns2016modular,kim1995architecture,salespardo2007extracting,rodriguez2017optimal}. The pervasive presence of these properties across such disparate systems stands in contrast to the divergent forces that shape them. Very large scale integrated (VLSI) circuits, for example, must balance logic capability with physical wiring length \cite{greenfield2007implications}. Similarly, natural neural networks face trade-offs between metabolic costs of wiring and effective transmission of information \cite{bullmore2012economy}, particularly over the long distances separating cortical and subcortical processing units in the human brain \cite{bassett2010efficient}. Furthermore, biological distribution networks facilitate complex life-critical functions by transmitting nutrients across an organism (or part of an organism) \cite{chappell2012how}, but are limited by communication distance between cells or tissue volumes \cite{papadopoulos2016embedding}. Even spatially embedded distributions systems \cite{barthelemy2011spatial} that transport humans across the city of London face trade-offs in movement efficiency and construction costs that favor modular architectures \cite{sperry2016rentian}. These competing constraints, while different for different systems, can be similarly arbitrated by fractal-like hierarchically modular structure \cite{ravasz2003hierarchical}, where a number of information processing units -- or \textit{modules} -- are recursively subdivided into smaller and smaller modules \cite{bassett2013multiscale}. \\

Scientific literature faces similar competing constraints, and it is intuitively plausible that its architecture might display similar topological principles. Intuitively, to be effective at transmitting information, thoughts must be organized into local co-relations among words \cite{alvarez2006hierarchical}, as well as more long-range co-relations among ideas across sentences and eventually across paragraphs, sections, chapters, and volumes \cite{schenkel1993long,amit1994language}. One could intuitively hypothesize, then, that the organization of a scientific research paper might be naturally and profitably studied with hierarchical network models \cite{i2001small}, similar to models used to study the hierarchal structures seen in both natural (e.g. neural information systems \cite{rubinov2011neurobiologically}) and engineered (e.g. the World Wide Web \cite{ravasz2003hierarchical}) systems. Such an effort could complement prior efforts in network modeling of scientific literature focused on understanding connections through citations \cite{kajikawa2007creating} or technical terms and topics \cite{hoffmann2004gene,rzhetsky2015choosing}.  One could further hypothesize that the organization of scientific literature changes as a given exposition is polished, revised, reframed, and reformulated during the internal writing process, the peer-review process, and the publication process. Understanding exactly how the organization of scientific literature and the ideas contained within it evolve with careful revision, could provide important insights into how that organization can be fine-tuned to maximize impact on the research community.  Such an effort could also complement prior work in modeling the temporal emergence of semantic networks in children \cite{hills2009longitudinal, steyvers2005large, utsumi2015complex, sizemore2017knowledge} by yielding new insights into the evolution of structure in semantic networks explicitly generated and used by adults with the goal of transmitting ideas.\\ 

Motivated by these open questions and hypotheses, we sought to determine whether the architecture of semantic networks built from scientific manuscripts was consistent with hierarchical modularity, and further we sought to understand how that hierarchical organization evolved through multiple drafts and revisions in support of optimal information transmission. We operationalized these goals by studying the existence and temporal variation of Rentian scaling \cite{bassett2013multiscale}, a notion of complexity in hierarchical, fractal-like structures. E. F. Rent made an observation in the 1960s, then reported by Landman and Russo in 1971 \cite{landman1971pin}, that the number of terminals ($T$) in the logic blocks of an integrated circuit scaled with the number of gates ($g$) within the blocks according to a power law: $T = tg^\beta$, where $t$ and $\beta$ are constants \cite{christie2000interpretation,landman1971pin}. In particular, $\beta$ is referred to as the Rentian scaling exponent, and has been shown to describe aspects of fractal network design \cite{greenfield2007implications} measuring the hierarchical modularity of a system \cite{bassett2010efficient}. However, Rent's rule is not limited to characterizing circuit design but extends well into systems biology and technology \cite{reda2009using}. Indeed, it has also proven useful in understanding the topological architecture of networked systems as diverse as the \textit{C. elegans} neuronal network \cite{bassett2010efficient}, the London Underground \cite{sperry2016rentian}, mycelial distribution networks \cite{papadopoulos2016embedding}, and the Internet \cite{cuesta2016exploring}, as well as in designing networks for neuromorphic systems \cite{partzsch2012developing,partzsch2015network}.  Here, we use this approach to understand how the Rentian scaling exponent changes throughout revisions of a scientific manuscript, influenced by tradeoffs in network complexity and efficiency. \\

\begin{figure}[h!]
	\centering
	\includegraphics[width=1.0\columnwidth]{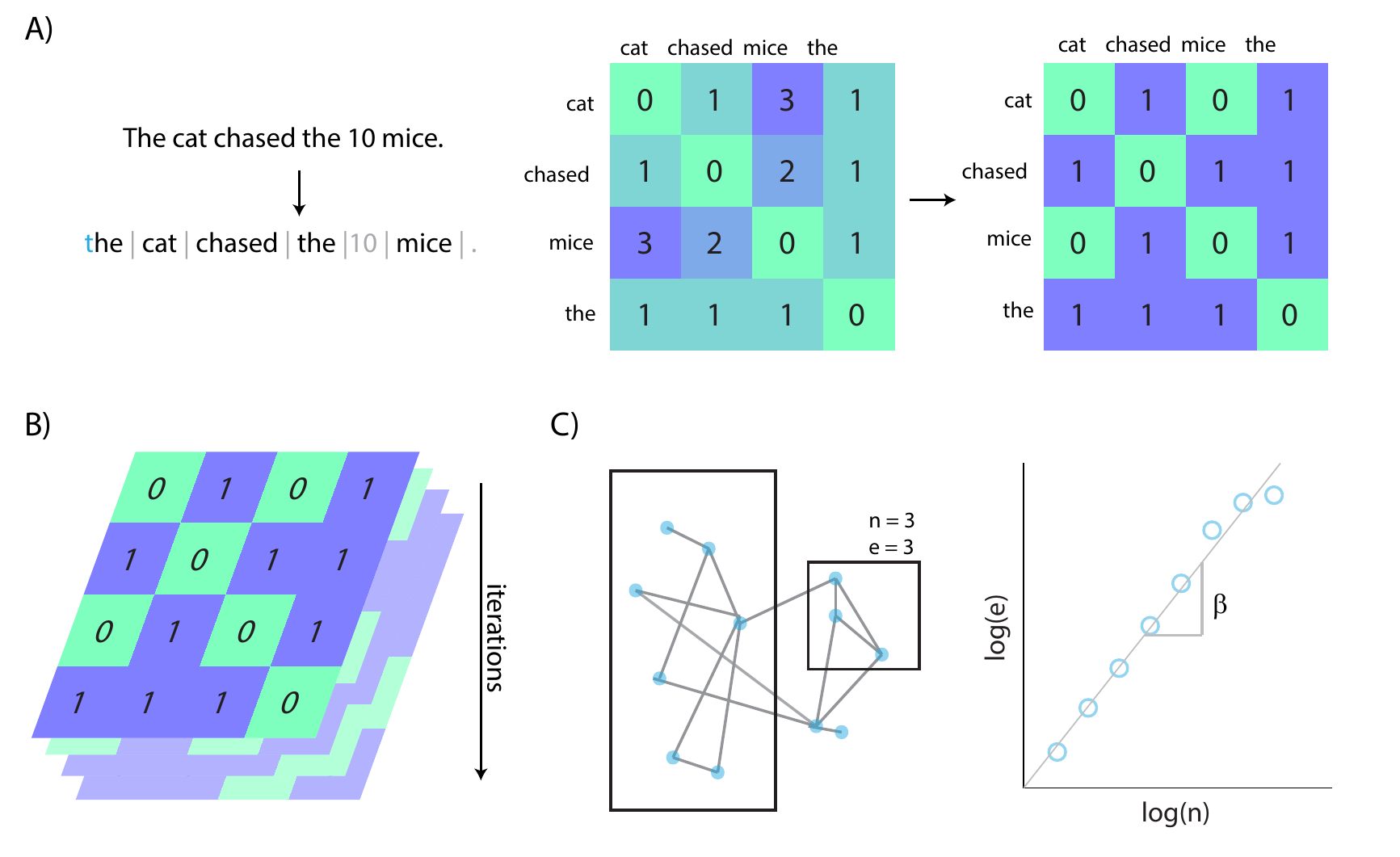}
	\caption{\textbf{Schematic Overview of the Approach.} \emph{(A)} Manuscript text is preprocessed by normalizing all characters (converting to lower-case) and tokenizing into discrete words. Next, tokens are filtered to remove numeric values and formatting commands embedded within the manuscripts. Here, we show the normalization and tokenization steps on a toy sentence \emph{(left)}. A network is defined in which each node is a unique word in the document, and the ij$^{th}$ edge is weighted according to the distance between word $i$ and word $j$ \emph{(middle)}. The network is thresholded into a binary matrix, in which non-zero values represent an edge between a pair of words that occurred within a threshold of $t$ words from each other. In this example, we use a threshold of $t=2$ \emph{(right)}. \emph{(B)} A network is constructed for each draft iteration of a paper to form a multilayer network, with nodes  aligned across layers. \emph{(C)} To compute the topological Rentian exponent, we first cover the entire graph with a single box, and then we recursively partition the box into halves to minimize the number of edges crossing the box boundaries. In each iteration, we count the number of nodes within the box, and the number of edges crossing the boundaries of the box. Rent's exponent is the slope of the linear relationship between the logarithm of the number of nodes and the logarithm of the number of crossing edges. We computed the exponent for each layer of the multilayer matrix, for each manuscript.
		\label{fig1}}
\end{figure}

To this end, we model each of 32 scientific manuscripts -- containing between three and 26 iterations of revision -- as a multilayer network, in which each layer reflects a different revision of that manuscript \cite{kivela2014multilayer}. In each network, nodes represent unique words, and an edge connects two nodes if the two words represented by those two nodes appear within the same 5-word window \cite{smadja1993retrieving}. We restrict ourselves to an examination of the introduction section of each manuscript, because it contains a diverse range of semantic structures and tends to undergo heavy revisions during the writing process. We compute the Rentian scaling exponent for each iteration of each manuscript to study the scaling trends over the manuscript's revision lifetime (see Figure \ref{fig1}). To preview results, we found that the scaling trends of the manuscripts over revision iterations differ from what would be expected in appropriate random network null models. Moreover, the manuscripts clustered into three main categories of those whose Rent exponent rose with revision, fell with revision, or remained relatively stable with revision, consistent with variations in the balance between network efficiency and network complexity in the semantic networks. Finally, we observed a negative correlation between the scaling exponent of the final manuscript revision and the number of authors involved in the manuscript, suggesting an important role for the number of idea generators (and perhaps the social structure between them) in the final organization of the semantic networks.\\

%and in the contextual variance of words in the manuscripts to their specificity to the article topic.

\section{Methods}

\subsection{Data Collation and Preprocessing}

We collated a novel database of $m=32$ manuscripts formatted in \LaTeX~from the Complex Systems Group in the Department of Bioengineering at the University of Pennsylvania. For additional information on the manuscripts, including final publication data,  the type of authors (undergraduate, graduate student, postdoctoral fellow, faculty member), and sex of authors, see the Supplementary Data table. Broadly, the manuscripts spanned the topics of computational neuroscience, bioengineering, network science, and complex systems. Each manuscript underwent between three and 26 iterations of revision (mean: 12.53; standard deviation: 5.81). We also collated both the number of authors on each manuscript, and the impact factor of the journal where the paper was published, to be used as covariates of interest in later analyses.

We began by extracting the introduction from each document. Early iterations of the manuscripts with incomplete or missing introductions were excluded from analyses; a total of 16 versions across 5 manuscripts were excluded. We then normalized all text by converting to lower case. Next, we used the Python Natural Language Toolkit (NLTK, version 3.1) to tokenize the text into word tokens, and the words were filtered to exclude formatting commands and punctuation. Specifically, we excluded \LaTeX\ comments, text embedded in math mode, numerical tokens, citations, and references to figures and tables. We also removed control commands (e.g., bold font, italicization, or colored text), although the text between these commands was maintained. For additional analyses that consider the effects on network structure of the morphological variants of the English words observed, see the Supplement.

% Omitted docs:
% AK_structure_learning_revisions: v1.1, v1, v2, v3_AK, v10_DBAKJV, v10_DBAKJVAK, v11_neuron
% CG_gran_mats: gran_mats_topology_arxiv
% CG_twos_company: arxiv, v16_review_early
% DB_autonomy_NN: v1, word_v14, word_v15, word_v16, word_v17
% ZZ_wavelet_paper_archive: v1
	
\subsection{Network and Multilayer Network Construction}

Next, we constructed a multilayer network \cite{kivela2014multilayer} for each of $m$ manuscripts. Each layer $l$ of the network corresponded to an iteration of the manuscript; thus, the multilayer networks contained between three and 26 layers ($3 \leq l \leq 26$). The number of nodes $n$ in each network was the vocabulary size of the manuscript; that is, the number of unique words appearing in the manuscript. Each node corresponded to a word in the manuscript, and nodes were aligned across layers, such that node $i$ in layer $l$ referred to the same word as node $i$ in layer $r$. For every pair of nodes $i$ and $j$, the edge between them was weighted in layer $l$ by the minimum number of words between words $i$ and $j$ in the $l$-th iteration of the manuscript. Note that this procedure created a nearly fully connected graph: a node in each layer was connected by a weighted edge to every other node in that layer, unless the word corresponding to that node did not appear in a certain iteration of the text. 

While the fully weighted network could be considered for some types of network analyses, the Rentian analyses that we describe here are theoretically understood and well-motivated only for binary networks. Thus, we next thresholded the adjacency matrix to retain an unweighted edge for words that appeared within 5 words of each other (i.e., there were fewer than five words between them) at least once in the $l$-th iteration. We chose the threshold based on lexicographic evidence that most lexical relations involve words that are within a neighborhood spanning five words \cite{smadja1993retrieving}. See the supplement for complementary results for neighborhoods spanning 3 words, and for neighborhoods spanning 7 words.  Thus, each manuscript is represented by an order 3 tensor $\mathbf{A}$, whose element $A_{ijl} = 1$ if word $i$ and word $j$ were five or fewer words apart in the $l$-th iteration of the manuscript. 

\subsection{Rentian Scaling}

In VLSI circuits, a simple power law known as Rent's rule has been shown to define the relationship between the number of processing elements (nodes) in a partition of the circuit and the number of external connections (edges) to that partition \cite{landman1971pin,christie2000interpretation}. Rent's rule can be described by the equation: 

\begin{equation}
E = kN^\beta
\end{equation}

where $E$ is the number of edges crossing partition boundaries, $N$ is the number of nodes within a block boundary, the constant $k$ is Rent's coefficient, and the constant $\beta$ is Rent's exponent. To estimate Rent's exponent in our multilayer networks, we first used a topological partitioning algorithm (hMetis software, version 1.5), which recursively partitions the network into halves, quarters, eighths, \emph{et. cetera}, following prior work \cite{klimm2014resolving}. For each partition, the average number of nodes within the partition and the average number of edges crossing the partition were computed \cite{cuesta2017method}. We estimated Rent's exponent $\beta$ as the slope in log-log space of the average number of nodes and the average number of edges. To confirm the goodness of fit, we also calculated the Pearson correlation coefficient between $log_{10}(E)$ and $log_{10}(N)$. For graphical visualizations of Rent's exponent and Pearson correlation coefficients for all manuscripts and all iterations, see the Supplement. \\

We estimated Rent's exponent individually for each layer of each multilayer network, where each layer can be viewed as a binary adjacency matrix with unweighted edges. Because the computation of Rent's exponent is nondeterministic  \cite{klimm2014resolving}, we estimated the scaling exponent a total of 100 times for each layer. We also performed a multiple linear regression to remove the effects of network sparsity, network density, and length of the text on the Rent's exponents (see Supplement), and all subsequent analyses were performed on the residuals. This approach allowed us to operationalize questions about the evolution of the textual network throughout the lifespan of the manuscripts by considering changes in Rent's exponent over manuscript iterations.\\

\subsection{Identifying Distinct Trajectories of Rent's Exponent}

We next constructed an $m \times m$ matrix $\mathbf{C}$ whose element $C_{ij}$ gave the Pearson correlation coefficient between (i) the time series of Rent's exponents for manuscript $i$ and (ii) the time series of Rent's exponents for manuscript $j$. We chose a Pearson correlation to assess general linear trends, due to simplicity of interpretation and the size of our dataset; assessments of nonlinear trends could be interesting if the dataset could be expanded significantly. We note that because these time series could range in length from 3 time points (corresponding to a multilayer network with $l=3$) to 26 time points (corresponding to a multilayer network with $l=26$), we first interpolated the time series of Rent's exponents for each manuscript using a linear interpolation, such that the time series length for each manuscript was equal to the maximum possible (26). (For graphical visualizations of these interpolated trends, see the Supplement.)\\ 

To algorithmically detect community structure in $\mathbf{C}$, we used a generalized Louvain-like locally greedy algorithm \cite{blondel2008fast} to maximize the following modularity quality function \cite{newman2006modularity}:
\begin{equation}
Q = \frac{1}{2\mu}\sum_{ij}\left(C_{ij}-\gamma \frac{k_i k_j}{2 \mu} \right) \delta(g_i,g_j)\,
\end{equation}
where $C_{ij}$ is a pairwise correlation (or \emph{edge}) between nodes $i$ and $j$, $k_i = \sum_j C_{ij} $ is the sum of weights attached to node $i$, and $g_i$ is the cluster to which node $i$ is assigned. The $\delta$ function is 1 if $g_i = g_j$ and zero otherwise, and $\mu = \frac{1}{2}\sum_{ij}C_{ij}$. The parameter $\gamma$ is a structural resolution parameter that can be used to tune the number of communities identified. Following prior work, we set $\gamma$ to the default value of $1$ \cite{bassett2011dynamic}. Due to the stochastic nature of the algorithm \cite{good2010performance}, it is not guaranteed to reach the global maximum modularity $\mathcal{Q}$. Thus, we applied the algorithm 100 times to identify persistent communities in a representative partition.\\

Broadly speaking, this community detection approach aims to cluster network nodes into densely interconnected groups called \emph{communities} or \emph{modules}, and has correlaries in maximum likelihood methods for community detection \cite{newman2016equivalence}. A higher value of the modularity $\mathcal{Q}$ indicates a better partitioning of the nodes into communities of similar Rentian scaling trends over the revision process. The resulting clusters from the Louvain algorithm display more similar scaling trends within each cluster, and less similar scaling trends between clusters. Manuscripts that are more strongly correlated in activity belong to the same module, while manuscripts that are less correlated in activity belong to different modules.  \\

\subsection{Null Models}

We next sought to determine whether the scaling trends we observed in the real data were different from those expected in appropriate random network null models. We considered two distinct null models. The first null model -- which we termed the \textit{statically rewired} null model -- probes the significance of the scaling exponent within each individual draft of a given manuscript. Following prior work \cite{bassett2011dynamic}, we rewired the edges within each layer of the multilayer network uniformly at random, such that the total number of edges within a layer remains the same, but each edge in the null model randomly connects an arbitrary pair of nodes. Intuitively, this model then represents the scaling exponent if the relationships between words were structured randomly within each iteration. Over all manuscripts, we computed 10 such rewirings for each layer of the multilayer network, and estimated the Rentian scaling exponent for each rewiring 10 times due to the nondeterministic nature of the exponent calculation. For graphical visualizations of the Rentian exponents for the statically rewired null model, see the Supplement. \\

The second null model -- which we termed the \textit{dynamically rewired} null model -- probes the significance of the change in scaling exponent over manuscript iterations. Here, we rewired the edges of the multilayer network uniformly at random across all layers (for additional considerations in choosing this and similar dynamic network null models, see \cite{sizemore2017dynamic}). Therefore, the node identities are preserved and the number of edges throughout the entire multilayer network remain the same, but each edge in the null model randomly connects an arbitrary pair of nodes in an arbitrary layer of the network. Intuitively, this model then represents the scaling exponent if the relationships between words were structured randomly across all iterations. Similar to the statically rewired null model, we computed 10 of such rewirings for each multilayer network, and estimated the Rentian scaling exponent 10 times for each layer of the network. For graphical visualizations of the Rentian exponents for the dynamically rewired null model, see the Supplement. 

To compare the true Rentian scaling trends to the scaling trends of the null graphs, we used techniques from a branch of statistics known as functional data analysis \cite{ramsay2006functional}. We first computed the area between the two scaling trends by subtracting the mean of the null graph scaling exponent from the mean of true scaling exponent for each iteration, and then we summed the absolute value of these differences over all iterations (in a manner similar to that described here \cite{bassett2012altered}). Specifically, we calculated the statistic $\zeta = \sum_{i=1}^T {| \bar{\beta_i}} - \bar{\beta_{i, null}}| $ where $T$ is the total number of iterations, $\bar{\beta_i}$ is the mean exponent at iteration $i$ because exponents were simulated 100 times. Next, we permuted the values between the null scaling exponents and the true scaling exponents, and computed the area, $\zeta$, from these permutations to obtain $\zeta_{\mathrm{perm}}$. Finally, we compared the true area, $\zeta$, to the null distribution, $\zeta_{\mathrm{perm}}$, to obtain an estimated $p$-value.\\

\subsection{Contextual Flexibility}

To measure contextual flexibility for each word in the manuscript, we computed the number of different words appearing within a five word window before or after the target word. However, because the size of the context set is largely driven by the frequency of the target word, we studied context size in relation to the number of word appearances (i.e., the size of the context set divided by the word frequency in the manuscript). We observed that the words with the lowest contextual flexibility corresponded to ideas most relevant to the topic of each manuscript. Finally, we considered only the words with the highest contextual flexibility; this required that we remove words that appeared five times or fewer across all iterations of the manuscript, as words that have few appearances have artificially high contextual flexibility due to their low frequency.

\section{Results}

\subsection{Existence of Scaling Behavior}

We first observed that all manuscripts displayed robust Rentian scaling, with Rent exponents greater than zero ($\beta>0$) for all manuscripts (see Supplement for all graphical visualizations of Rentian exponents for all manuscripts and all revision iterations). To confirm these results, we also calculated the Pearson correlation coefficient between $log_{10}(E)$ and $log_{10}(N)$, and found that $r > 0.99, p < 0.001$ across all manuscripts. We further noticed that Rent exponents varied from manuscript to manuscript, ranging from approximately $\beta=0.69$ to approximately $\beta=0.81$, consistent with the range of values reported elsewhere in neuronal networks and VLSIs and consistently greater than the range of values reported for RAM \cite{bassett2010efficient}. These results indicate that Rentian scaling is a robust property of these textual networks constructed from scientific research articles across revisions.

\subsection{Dynamics of Scaling Behavior}

\begin{figure}[h!]
	\centering
	\includegraphics[width=1.0\columnwidth]{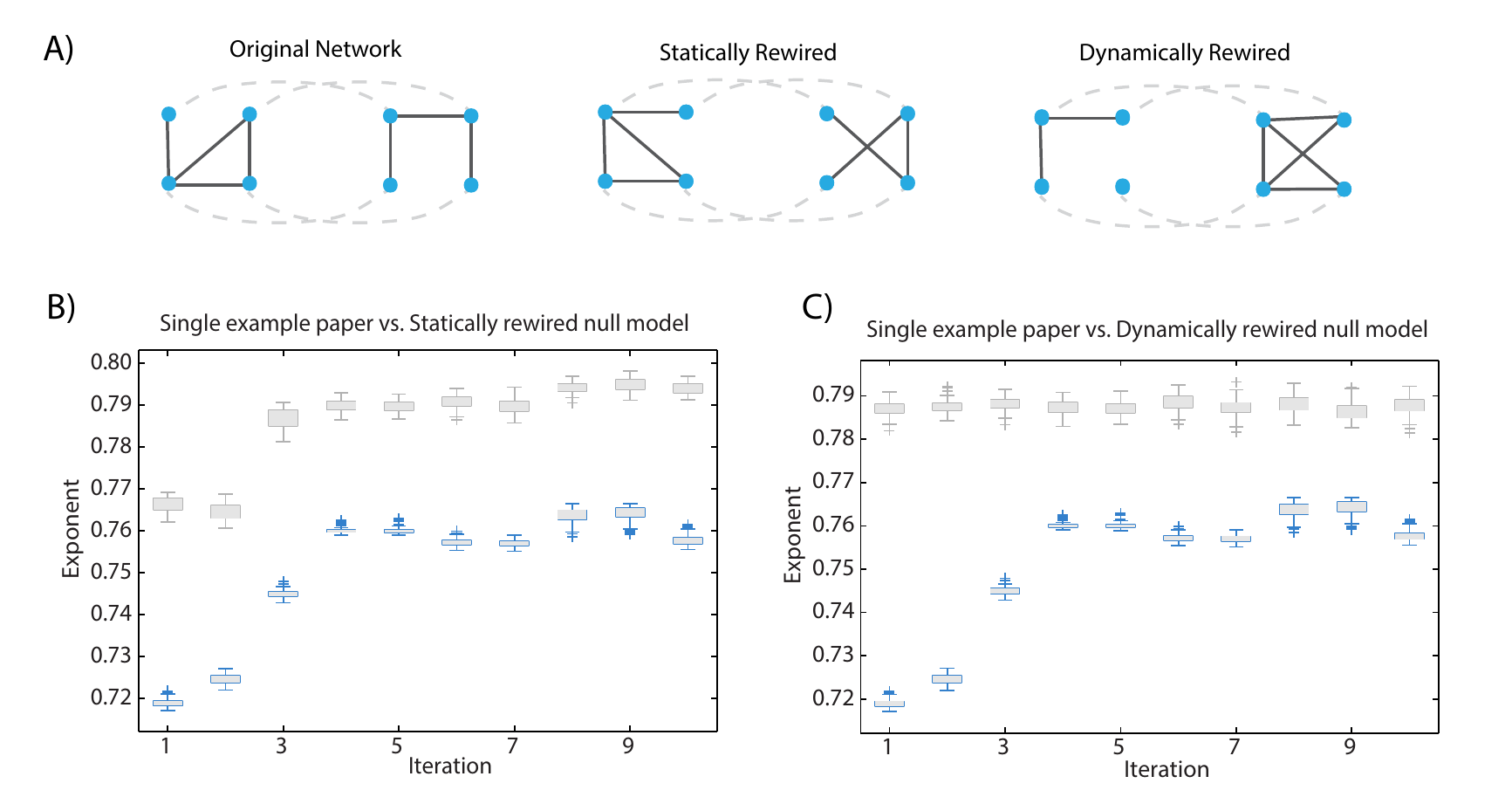}
	\caption{\textbf{Temporal Variations in Rentian Scaling: Comparison to Random Network Null Models.} \emph{(A)} A toy network of two layers and four nodes per layer. Dashed gray lines indicate alignment in node identity across layers. In the statically rewired null model, edges are rewired uniformly at random \emph{within} each layer. In the dynamically rewired null model, edges are rewired uniformly at random \emph{across} layers. \emph{(B)} Rent's exponent of the statically rewired null model (shown in gray) along with Rent's exponent of a representative manuscript over 10 iterations of revisions (shown in blue) for a representative manuscript. \emph{(C)} Rent's exponent of the dynamically rewired null model (shown in gray) along with Rent's exponent of a representative manuscript over 10 iterations of revisions (shown in blue) for a representative manuscript. For graphical visualizations of similar results across all manuscripts, see the Supplement.
		\label{fig2}}
\end{figure}

After conducting scaling analyses across the 32 manuscripts, we next sought to determine whether the temporal trends that we observed in the scaling exponent over revisions would be expected in appropriate random network null models (Figure \ref{fig2}A). Thus, we defined a statically rewired null model where we rewired edges uniformly at random within each layer of a multilayer network corresponding to a single manuscript. We then estimated Rentian scaling exponents on these null model networks. In comparing the Rentian scaling trends of the manuscripts to those observed in the random network null models, we observed that the two trends differed significantly (permutation test using functional data analysis: $p < 0.01$ for all manuscripts, see Methods). Here in the main text of this article, we show a representative plot of the null model for a single manuscript (see Figure \ref{fig2}B) and we show the remainder of the plots for other manuscripts in the Supplement. This finding indicates the presence of a salient structure in the textual networks, which is not maintained after random permutation of the edges. \\

To further and more directly probe the significance of the evolution of the textual networks, we constructed dynamically rewired null models by rewiring edges uniformly at random across layers. For each null model network, we again estimated Rentian scaling exponents. This process allowed us to examine the null hypothesis that the structure of the textual networks does not change over revisions. Again, we observed that the two trends differed significantly (permutation test using functional data analysis: $p < 0.01$ for all manuscripts, see Methods). Here in the main text of this article, we show a representative plot of the null model for a single manuscript (see Figure \ref{fig2}C) and we show the remainder of the plots for other manuscripts in the Supplement. This finding indicates that -- while the scaling exponent of the true networks varies over the iterations -- the exponent of the null model remains constant over time, suggesting that the change in the scaling exponent over manuscript revisions is not explained by random fluctuations in edge connectivity.\\

\subsection{Distinct Trends in Scaling}

\begin{figure}[h!]
	\centering
	\includegraphics[width=1.0\columnwidth]{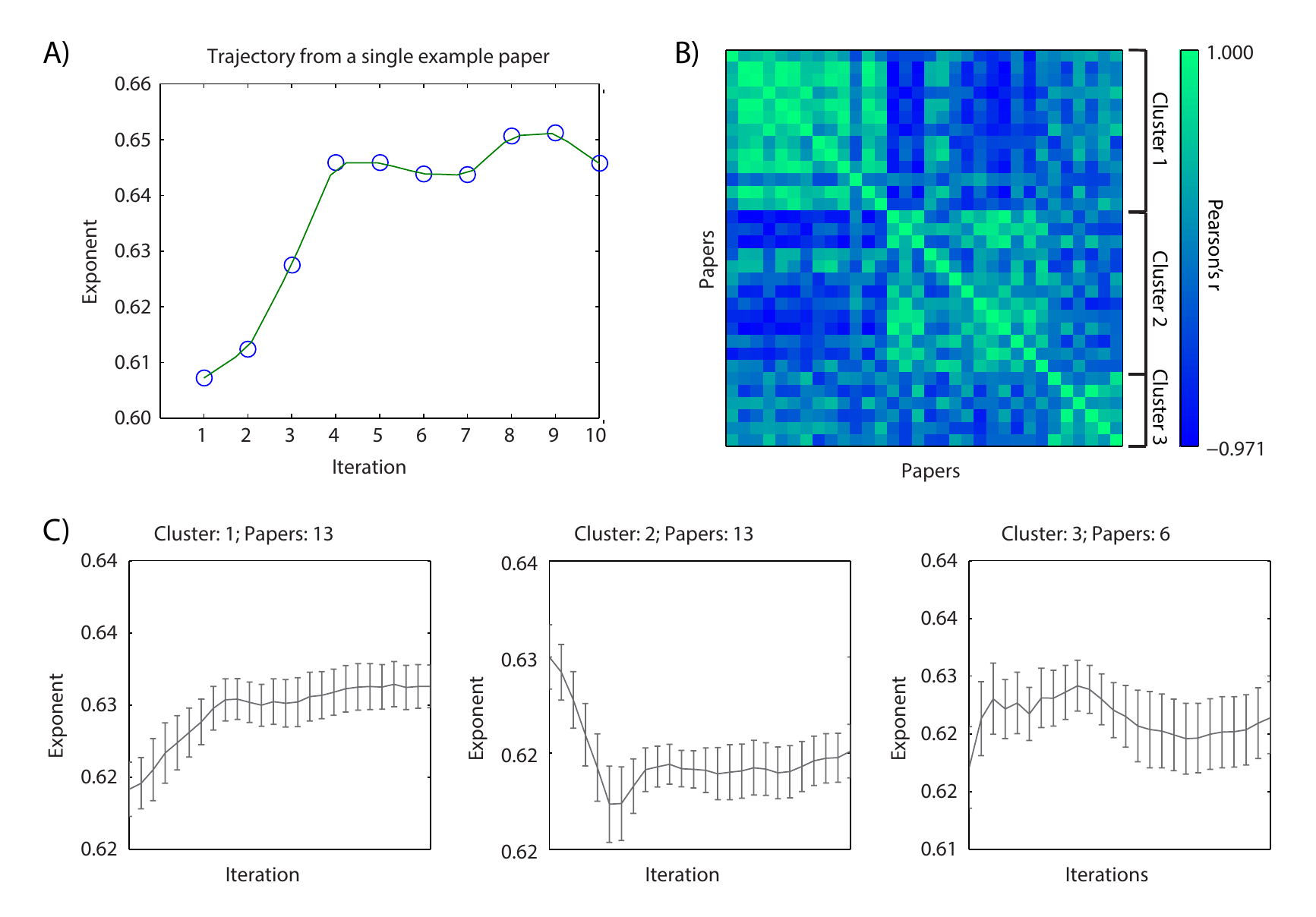}
	\caption{\textbf{Clustering of Scaling Trends.} \emph{(A)} The time series of Rentian exponents was interpolated for all manuscripts using a linear interpolation. The interpolation for a representative manuscript is shown in green, with the true scaling exponents shown in blue circles. \emph{(B)} Using the interpolated time series, we computed Pearson's correlation coefficient between the scaling trends of every pair of manuscripts, and we applied a community detection algorithm to the resultant correlation matrix to uncover three clusters of trends. \emph{(C)} We show the mean exponent and standard error over the interpolated iterations for each of the three clusters. %panel A shows Ankit's EpiNetNMF
		\label{fig3}}
\end{figure}

We next examined common trends in the scaling exponents throughout the iterative manuscript revision process. Namely, we sought to identify clusters of manuscripts that exhibited similar scaling trends over iterations. Because the manuscripts range between three and 26 rounds of revision, we interpolated the iteration count of each manuscript using a linear interpolation, such that the interpolated iteration count of each manuscript was equal to the maximal iteration count of all manuscripts (26 iterations, see Figure \ref{fig3}A). We computed the Pearson correlation coefficient between the interpolated scaling trends of every pair of manuscripts, and used a common community detection algorithm to identify community structure in this correlation matrix (see Methods and Figure \ref{fig3}B). We note that we used community detection in the form of modularity maximization rather than $k$-means clustering due to its incorporation of an explicit random network null model in the form of the configuration model. We observe that the modularity values obtained from the true data were significantly greater than the modularity values obtained from the null modal data ($t_{198} = 428.69$, $p<0.001$), suggesting an expectedly high level of meso-scale clustering in these data. \\

The community structure obtained from this approach was comprised of three separate modules. Because we identified these clusters by modularity maximization, the correlation between the scaling trends within clusters was higher than the correlation between scaling trends across different clusters. The first cluster encompassed 13 manuscripts whose scaling exponent broadly increased over revisions. The second cluster encompassed another 13 manuscripts whose scaling exponent broadly decreased over revisions. The third cluster encompassed the remaining six manuscripts whose scaling exponent remained relatively stable throughout the revision process (Figure \ref{fig3}C). We note that due to the relatively small size of our data set, we are not powered to describe nonlinear trajectories in the scaling exponents. The existence of these three trends suggests a tradeoff between the complexity of the network (which promotes an increasing scaling exponent) versus the efficiency in network wiring (which promotes a decreasing scaling exponent). 

\subsection{Relation Between Rentian Scaling, Impact Factor, and Authorship}

\begin{figure}[h!]
	\centering
	\includegraphics[width=0.65\columnwidth]{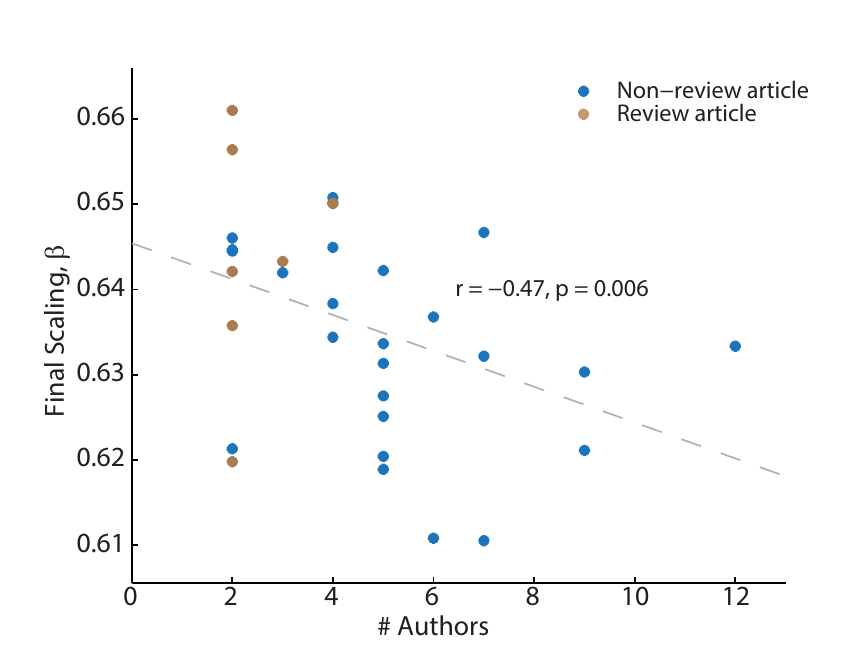}
	\caption{\textbf{Number of Authors is Related to the Rentian Scaling of the Final Manuscript.} We observed a relationship between the Rent's scaling exponent of the final iteration of the manuscript, and the number of authors associated with the manuscript. The scaling exponent decreases as the number of authors increases (Spearman rank correlation coefficient $r_s = -0.47, p = 0.0062$). Note that 2-author manuscripts tended to be review articles.
 		\label{fig4}}
\end{figure}

Next, we sought to address the question of whether and how the scaling exponents of the manuscripts might relate to tangible characteristics of the paper such as the impact factor of the journal in which the paper was published or the number of authors that contributed to the paper. After curating the impact factors of the journals at which these manuscripts were published, we found that there was no significant correlation between impact factor and final Rent's exponent ($p>0.05$). Of course, it is well-known that impact factor is a metric that is difficult to directly interpret, being heavily modulated by the field or subfields represented by the readership. However, we did find a relationship between the number of authors listed on the manuscript and the Rentian scaling exponent of the final iteration of the paper. Namely, as the number of authors associated with the manuscript increased, the final scaling exponent decreased (Spearman rank correlation coefficient $r_s = -0.47, p = 0.0062$; see Figure \ref{fig4}; note that 2-author manuscripts tended to be review articles, and note that we used a Spearman correlation due to the non-normality of the distribution). This relation is consistent with the notion that more authors is associated with greater efficiency in network wiring. The observed association between the number of the authors and the wiring efficiency in the textual networks suggests that the landscape of scientific writing is in part shaped by the social environment in which those discoveries take place. \\

\subsection{The Supportive Mechanism of Contextual Flexibility}
Finally, we sought to better understand how the words themselves, and the contexts in which they are used, could in theory produce the Rentian scaling observed in the textual networks. Intuitively, textual networks with hierarchically modular structure are likely to contain words that can be placed in one module at one level of the hierarchy, and another module at another level of the hierarchy. We refer to this phenomenon as \emph{contextual flexibility}. We studied the frequency and contextual flexibility of words in each manuscript after removing common stop words (very common words that impart little value in meaning, e.g., ``the'') using the NLTK English stopword list. We observed that the remaining words with the highest frequencies were pertinent to the major topics described in each manuscript (Table~\ref{Tab1}). Interestingly, the words with the highest contextual flexibility correspond to broader terms that were less specific to the topic of the paper, and could be more generally used in similar literature. These results provide a conceptual intuition for how Rentian scaling can occur in textual networks: words can be used in diverse contexts to create fractal, hierarchical structure, while words used in a narrower range of contexts provide a substrate for local modularity.

\begin{table}[]
	\centering
	\caption{\textbf{Contextual Flexibility.} We show the ten words with the highest and lowest contextual flexibility (contexts per count) in a representative manuscript. 
		%The remainder of the manuscripts are included in the supplementary material.
		}
	\label{Tab1}
	\begin{tabular}{|l|l|l|l|}
		\hline
		word          & contexts per count & word               & contexts per count  \\ \hline
		need          & 8                  & patients           & 1.1                \\
		\hline
		also          & 7.2                & spontaneous        & 1.1                \\
		\hline
		communication & 5                  & crippling          & 1.1                \\
		\hline
		time          & 4.71               & brain              & 1.09               \\
		\hline
		approaches    & 4                  & epochs             & 1.07               \\
		\hline
		two           & 3.67               & interictal         & 1                  \\
		\hline
		common        & 3.5                & electrocorticogram & 1                  \\
		\hline
		responsible   & 3.4                & recur              & 1                  \\
		\hline
		emerged       & 3.4                & linked             & 1                  \\
		\hline
		coincide      & 3.4                & million            & 0.7           \\
		\hline    
	\end{tabular}
\end{table}

% the numbers in this table come from AK_EpiNetNMF (same as the manuscript in fig 3)

\section{Discussion}

Similar to the hierarchical structure observed in many other natural and engineered systems, we have observed a comparable structure in the network embedding of the introductions of scientific manuscripts spanning research areas of network science, complex systems, systems biology, systems medicine, and neuroscience. Using the Rentian scaling exponent as a measure of the complexity of a fractal-like hierarchical system, these semantic networks evolve over the revision and publication cycle in a manner that diverges from that expected in appropriate random network null models. Furthermore, we observed the presence of three dominant clusters of scaling trends over manuscript iterations, suggesting competing forces of network complexity and network efficiency. Interestingly, the scaling exponent of a manuscript's final iteration is negatively related to the number of authors on the study, suggesting a dependence between the product of scientific collaboration and the social network implementing that collaboration. Finally, we noted a potential mechanism for Rentian scaling in the contextual flexibility of words in each manuscript, and found that words with lowest contextual flexibility corresponded to ideas most relevant to the topic of each manuscript while words with highest contextual flexibility corresponded to terms used ubiquitously across manuscripts.

\subsection{Hierarchical Modularity in Networked Systems}

Networked systems are present in a variety of contexts, spanning social science, urban planning, biology, and engineering. Despite disparate contexts, many of these networked systems share similarities in structure, reflecting common properties that promote information transfer and network adaptability \cite{barthelemy2011spatial}.  One critical property retained across many networked systems is hierarchical modularity \cite{simon1962architecture}, a fractal-like organization in which networks are divided into highly interconnected modules, each of which in turn are further subdivided into smaller and smaller modules \cite{song2005self}. Evolutionarily, this modular structure is advantageous in conferring adaptability to the system, making it possible to adapt a single module without changing the remainder of the system \cite{bassett2013multiscale}. Many real-world systems possess a natural hierarchy in structure -- for example in the committee network of the House of Representatives, in which the House floor is subdivided into groups of committees and further divided in groups of subcommittees \cite{mucha2010community}. Similarly, in the networks governing metabolism in \textit{E. coli} bacterium \cite{ravasz2002hierarchical}, co-actors in movies, and synonyms of words \cite{ravasz2003hierarchical}, a fractal structure in which the network clustering coefficent scales with the degree of a node has been observed. Here, we applied Rent's rule as a heuristic to quantify the hierarchical organization of a networked system.  Rentian scaling principles have been extensively studied in VLSI circuits \cite{landman1971pin,christie2000interpretation,bassett2010efficient}; however, in recent years, the same scaling principles have been shown to apply to a number of biological and physical systems such as the London Tube \cite{sperry2016rentian}, rodent vasculature and mycelial networks \cite{papadopoulos2016embedding}, and the human brain \cite{bassett2010efficient}. Here, we show a similar scaling trend in semantic networks of scientific articles, and moreover study how the scaling trend changes over the manuscript writing process. We observe that the scaling exponent does not remain constant over iterations of the manuscript, which we probed using statically rewired and dynamically rewired null models, reflecting an evolution of complexity throughout the writing process. This dynamic approach can be readily translated into the context of other networked systems, and may yield further insights regarding the growth and expansion of networks in biological, social, or engineering contexts over time.

\subsection{Constraints and Trade-offs in Network Organization}

While many systems exhibit a fractal-like, hierarchically modular structure, these systems are formed under the trade-offs of transport efficiency versus cost efficiency. As an intuitive example, consider the network formed by a city transportation system, in which nodes are transport stations and edges are routes taken between stations. While it is most efficient to have a direct route between every pair of stations, thus reducing the need for transfers, inevitably the cost of maintaining such an infrastructure is impractical to maintain. Thus, transport systems have evolved to balance the cost of maintaining routes with the cost of efficient travel across the system. In biology, studies have shown that networked systems are cost-efficiently (but not cost minimally) embedded in physical space \cite{bassett2010efficient,papadopoulos2016embedding}. Furthermore, these systems have evolved to prioritize varying constraints: rodent vasculature is optimized for low-cost wiring, while mycelial networks form more complex but expensive networks \cite{papadopoulos2016embedding}. These discrepancies are thought to reflect differing environmental pressures: while vasculature is wired to supply blood to a fixed region with highly regulated oxygen and molecular content, mycelial networks must be able to adapt to environmental fluctuations and local variations in non-uniform soil quality \cite{papadopoulos2016embedding}. In the human brain, a higher Rent's exponent in the cortex, compared to the cerebellum, when embedded in three dimensional space is consistent with the need for increased logical capacity of cortical systems \cite{bassett2010efficient}. A similar trend has been observed in computer chips, in which the Rentian scaling exponent in high performance computers exceeds that of simpler dynamic memory chips, suggesting the competing forces of logical complexity and wiring efficiency \cite{bassett2010efficient}. In our semantic networks, we observed three distinct clusters of scaling trends over time, yielding an increasing exponent, decreasing exponent, or relatively stable exponent over manuscript iterations. These results suggest that network complexity and efficiency must be balanced in semantic networks, and no one trajectory clearly dominates. Similar to the trade-offs faced by biological and electronic networks, our results suggest that embeddings of ideas into the fixed structure of a scientific research article faces comparable constraints in which both complexity and efficiency are taken into account.

\subsection{Network of Science within a Social Environment}

In recent years, there has been an increasing interest in the network of scientific discovery itself, and how this network can be configured to support advancements in science \cite{ke2014tie,bae2016multi}. Building the scientific network of chemical relationships in biomedicine using nodes as scientific concepts and edges as relations between these concepts, Rzhetsky et al. formulated a generative model of scientific discovery \cite{rzhetsky2015choosing}. This approach illustrated that the current scientific environment reinforces the exploitation of well studied chemicals, a conservative strategy which supports a career requiring steady research output. However, a more efficient strategy for network discovery entails linking items distant from each other in the network, an approach which involves considerable risk for experimental success \cite{rzhetsky2015choosing}. In our work, we identified an interesting trend between the number of authors associated with a manuscript and the Rentian scaling exponent of the final revision of the manuscript, namely that the exponent decreases as the number of authors increases. This trend may indicate a sort of consensus among authors, in which network complexity is reduced with an increasing number of writers, suggesting that scientific discovery is intertwined within a social environment which facilitates discovery. As social networks form a medium for the spread of information or ideas, in which network heterogeneity reflects interactions among diverse groups allowing individuals access to a larger selection of non-redundant social resources \cite{lewis2008tastes}, it is plausible that the social environment is also a driver in shaping the process of discoveries in science. With recent advancement in studying network configurations in both social interactions and scientific discoveries, we imagine that viewing the scientific environment embedded within a social landscape may yield fascinating insights into how social forces influence scientific progress.

\subsection{Methodological Considerations and Future Directions}

A number of methodological considerations are pertinent in evaluating the results of this study. Firstly, we collated iterations of revision for 32 manuscripts but we acknowledge that the size of this dataset is limited, constrained by the number of manuscripts for which intermediate draft iterations were readily available. We focused on the introductions of the manuscripts, because it contains a diverse range of semantic structures and tends to undergo heavy revisions during the writing process. We separated by word boundaries, omitting formatting commands and citations present in the raw form of the documents. This step is imperfect, and subject to typographical errors in the iterations of each manuscript. As such, errors made in formatting or spelling, for example if formatting commands or comments are not properly annotated, may impact the appropriate removal of these non-textual elements within the drafts of manuscripts. A further limitation of this approach is that although two manuscripts may have undergone the same number of revision iterations, the time between each round of revision could differ between the two. We sought to mitigate this issue by studying trends in the scaling exponent throughout the duration of the revision process.\\

In future, the analytical approach that we use here could be extended to other sections of the manuscript, enabling one to study whether Rentian scaling trends vary by section. It would also be fascinating to study scaling trends throughout the career of scientific researchers, using manuscripts published throughout a scientist's career. As these articles are publicly accessible, the size of the available dataset is orders of magnitude larger than the 32 manuscripts used here. Furthermore, as these published articles are final iterations of manuscripts, they would be less susceptible to the formatting or typographical errors present in intermediate drafts of manuscripts. This approach may also yield interesting insights into how scaling properties evolve over a scientist's career to promote consistent advancements in scientific discovery.

\section{Conclusion}

A modular hierarchical architecture is observed across various real-life networks. Here, we quantify this fractal-like structure in semantic networks composed of introductions of scientific articles, in particular throughout the drafting and revision process of scientific manuscripts. We observe that the Rentian scaling exponent describing hierarchical network structure varies throughout the publication life cycle and clusters into three main trends among our collection of manuscripts. This evolution of the scaling exponent over the manuscript revision process suggests a balance of network complexity and efficiency, a trade-off which similarly governs other natural and man-made networked systems. Our study lays the groundwork for future investigations into how such semantic network architectures might differ across disciplines, or across different sorts of papers (from incremental to pioneering), and the features that better reflect the paper's future impact versus the factors that better reflect the paper's generative social milieu.

\section{Acknowledgments}

We acknowledge helpful discussions with Florian Klimm, Lia Papadopoulos, Lyle Ungar, Mark Libermann, and Ani Nenkova. We also thank Ann E. Sizemore, Jordan Dworkin, and David Lydon-Staley for helpful comments on an earlier version of this manuscript. The work was supported by the John D. and Catherine T. MacArthur Foundation, the ISI Foundation, the Alfred P. Sloan Foundation, and NSF CAREER award PHY-1554488. The content is solely the responsibility of the authors and does not necessarily represent the official views of any of the funding agencies.

\singlespacing

\newpage
\bibliographystyle{ieeetr}
\bibliography{./bibfile}

\end{document}

\title{Supplementary Materials for\\ ``Evolution of semantic networks in biomedical texts''}
\author[1,2]{Lucy R. Chai}
\author[1,3,4,5]{Danielle S. Bassett}
\affil[1]{Department of Bioengineering, University of Pennsylvania, Philadelphia, PA 19104 USA}
\affil[2]{Program in Machine Learning,Speech and Language Technology, Department of Engineering, University of Cambridge, Cambridge, UK CB2 1ST}
\affil[3]{Department of Electrical \& Systems Engineering, University of Pennsylvania, Philadelphia, PA 19104 USA}
\affil[4]{Department of Neurology, University of Pennsylvania, Philadelphia, PA 19104 USA}
\affil[5]{To whom correspondence should be addressed: dsb@seas.upenn.edu}

\maketitle

\newpage
\section{Supplementary Results}

\subsection{Estimation of Rentian Scaling Exponent}

\begin{landscape}
\begin{figure}[ht!]
	\centering
	\includegraphics[width=1.0\columnwidth]{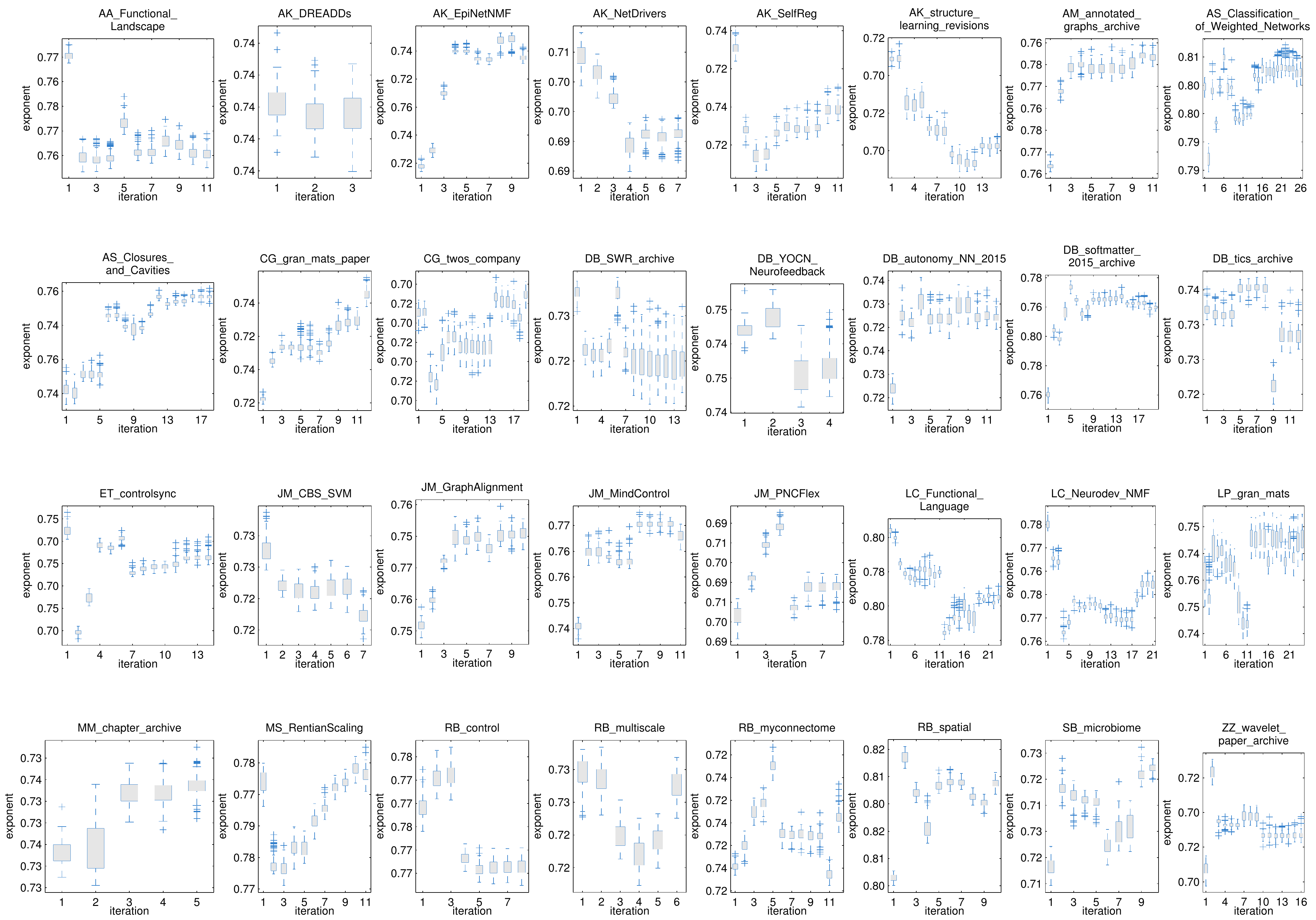}
	\caption{\textbf{Raw Scaling Exponents.} Rentian scaling trends for all manuscripts, without regressing out the effects of sparsity, density, and length of the texts.
		\label{s1}}
\end{figure}
\end{landscape}

\begin{landscape}
	\begin{figure}[ht!]
		\centering
		\includegraphics[width=1.0\columnwidth]{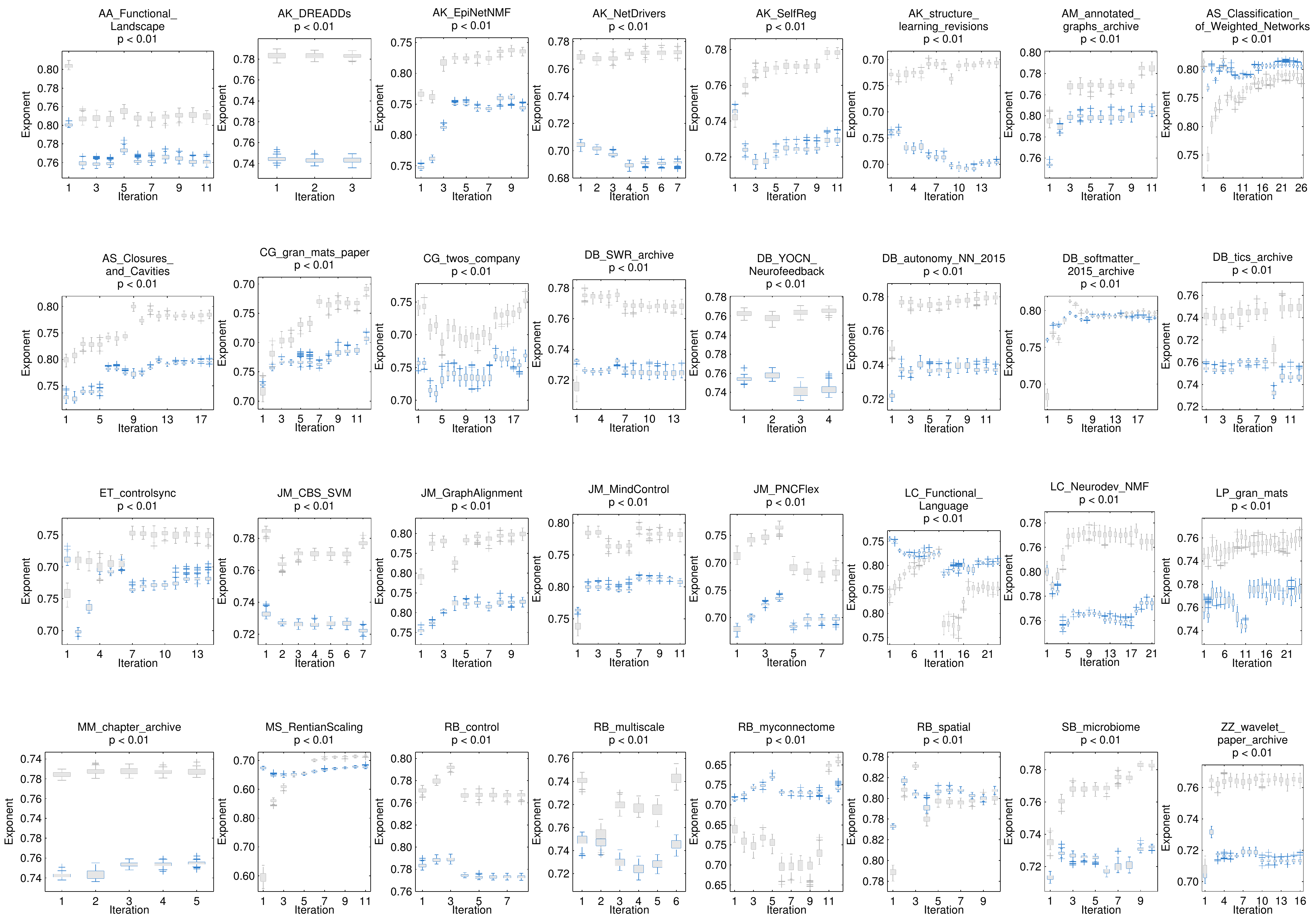}
		\caption{\textbf{Statically Rewired Null Model.}  Rentian scaling trends for all manuscripts (shown in blue), in comparison to the statically rewired network null models (shown in gray).
			\label{s2}}
	\end{figure}
\end{landscape}

\begin{landscape}
	\begin{figure}[ht!]
		\centering
		\includegraphics[width=1.0\columnwidth]{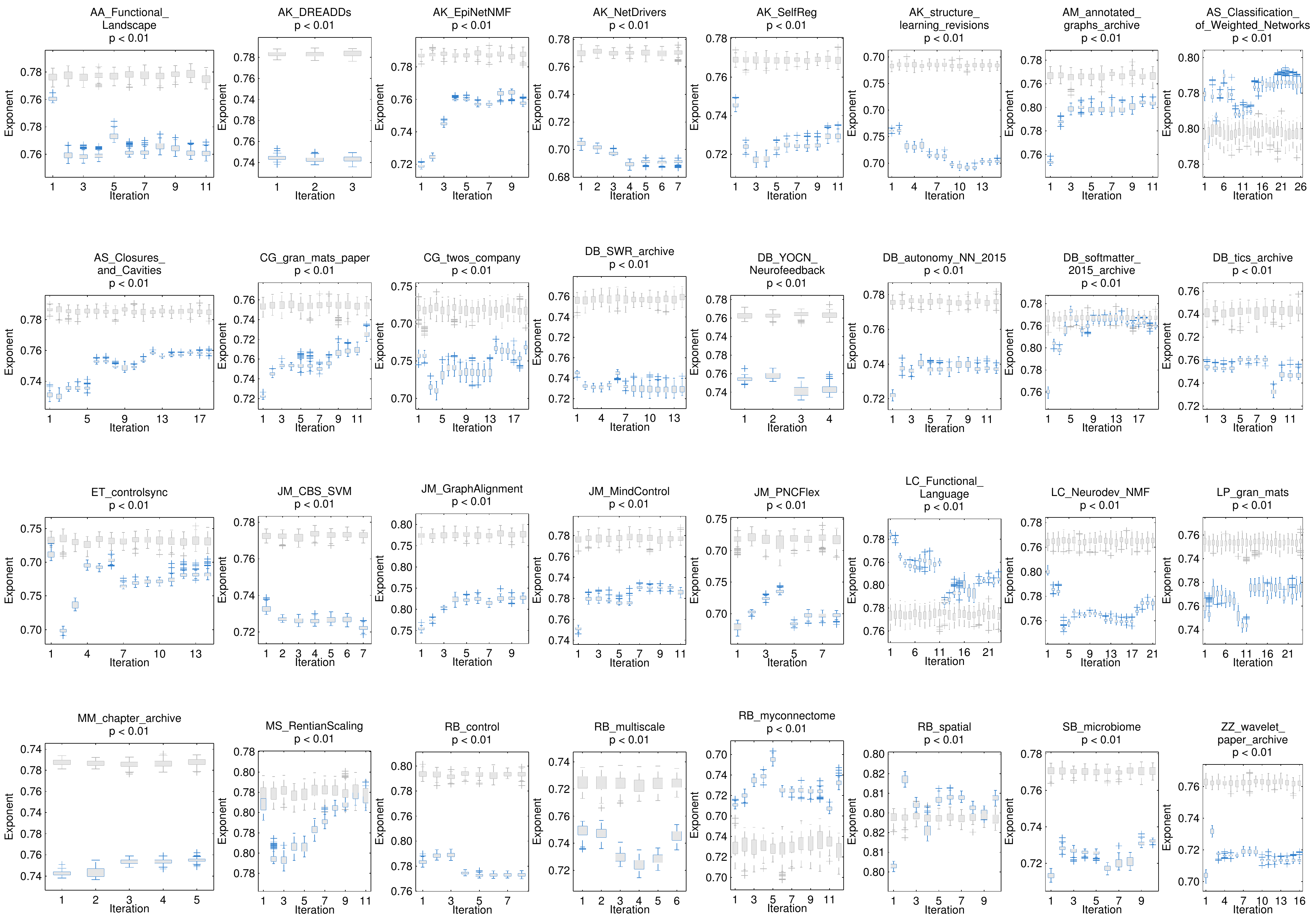}
		\caption{\textbf{Dynamically Rewired Null Model.} Rentian scaling trends for all manuscripts (shown in blue), in comparison to the dynamically rewired network null models (shown in gray).
			\label{s3}}
	\end{figure}
\end{landscape}

The power law governing Rent's rule is described by the equation $E = kN^\beta$ where $E$ is the number of edges crossing the boundaries of a network partition, $N$ is the number of nodes within the partition boundaries, the constant $k$ is Rent's coefficient, and the constant $\beta$ is Rent's exponent. To estimate Rent's exponent, we determined the slope of the linear least squares fit between $log_{10}(E)$ and $log_{10}(N)$. Across all manuscripts, we obtained a Pearson correlation coefficient of $r > 0.99, p < 0.001$ between $log_{10}(E)$ and $log_{10}(N)$. In Figure \ref{s1} we show the Rentian scaling exponents determined from this linear fit over all revision iterations of the 32 manuscripts. We next computed statically and dynamically rewired network null models, and we compared these scaling exponents to the true scaling exponents using nonparametric permutation testing approaches from a branch of statistics known as functional data analysis (see Methods). The null models and the true exponents for all manuscripts are shown in Figures \ref{s2} and \ref{s3}. \\

\subsection{Controlling for Sparsity, Density, and Length}
\begin{figure}[ht!]
	\centering
	\includegraphics[width=1.0\columnwidth]{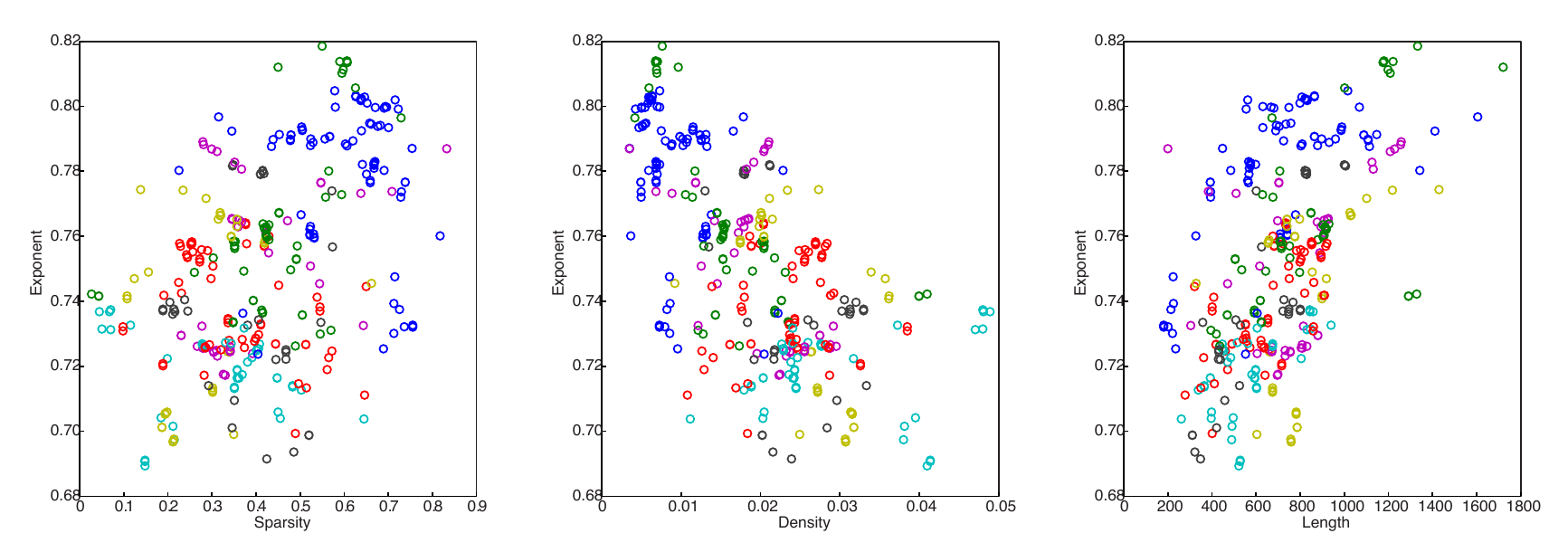}
	\caption{\textbf{Regression of confounding variables.} We observed that the sparsity (related to the vocabulary size), density (related to threshold value) and word count of the revisions of the manuscript were correlated with the scaling exponent. Thus, in all further analyses, we removed the effects of these variables by considering the residuals of a multilinear regression.
		\label{s4}}
\end{figure}

\begin{landscape}
	\begin{figure}[ht!]
		\centering
		\includegraphics[width=1.0\columnwidth]{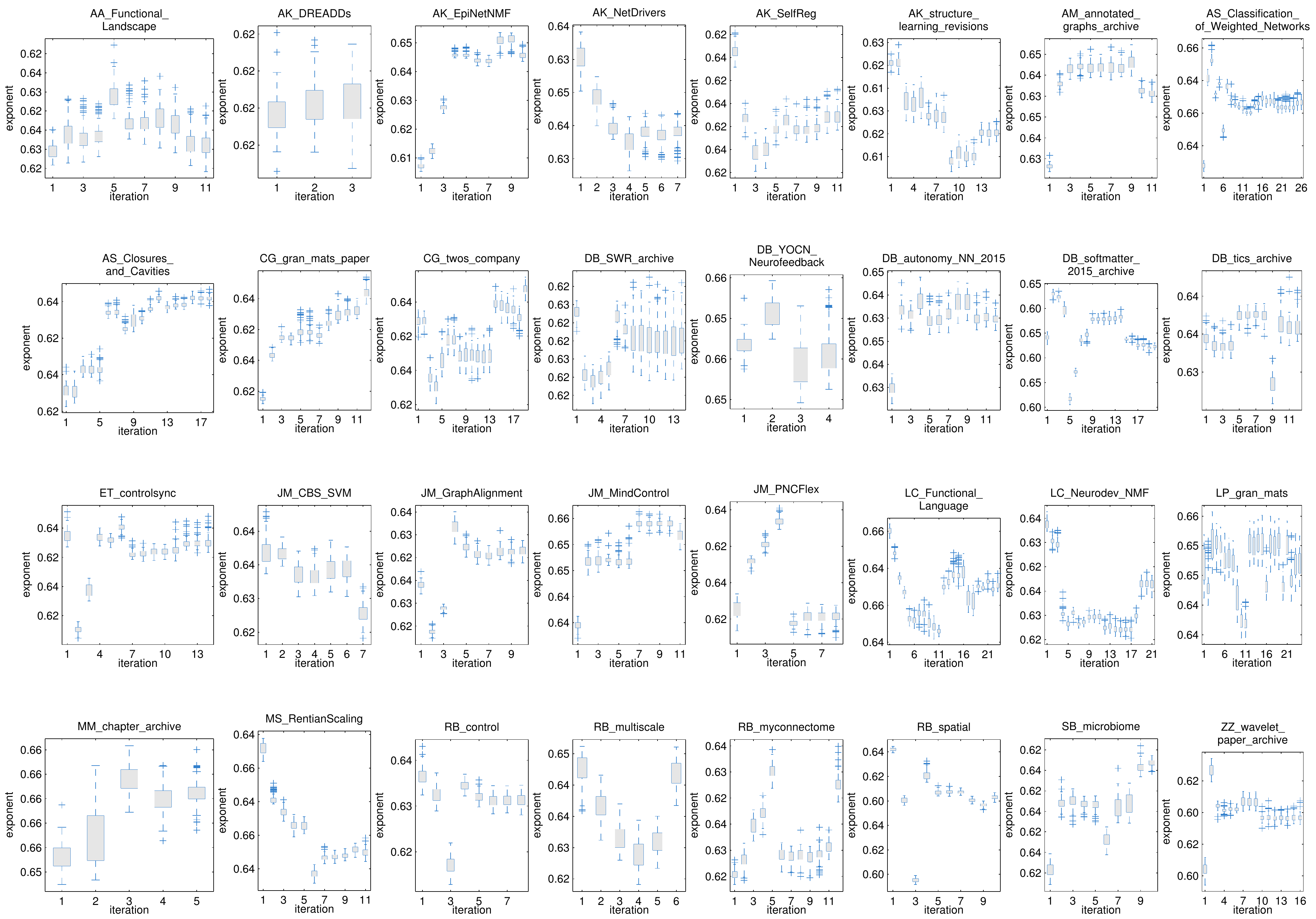}
		\caption{\textbf{Scaling Exponents after Regression.} Rentian scaling trends for all manuscripts, after regressing out the effects of sparsity, density, and length of the texts.
			\label{s5}}
	\end{figure}
\end{landscape}

As we sought to compare the scaling exponent across different manuscripts and iterations of revisions, we noticed that the scaling exponent was influenced by the sparsity (number of nodes) and density (number of edges) of the network, as well as by the length of the text (Figure \ref{s4}). When we considered each manuscript separately, we observed that 22 manuscripts displayed a negative correlation between scaling exponent and sparisty, and 10 manuscripts displayed a positive correlation between scaling exponent and sparsity. The trends became clearer when we considered all manuscripts and all interations of revisions together. Here, we observed significant correlations between the scaling exponent and all three variables (Sparsity: Pearson correlation coefficient $r = 0.46, p < 0.001$, Density: Pearson correlation coefficient $r = -0.65,p < 0.001$, Length: Pearson correlation coefficient $r = 0.56, p < 0.001$). We therefore performed a multiple linear regression to remove the effects of these variables, such that the scaling exponent after regression $\beta '$ is related to the original scaling exponent $\beta$ via the equation $\beta ' = \beta - c_1s - c_2d - c_3 l$ where $s$, $d$, and $l$ refer to sparsity, density, and length, respectively, and where $c_1$, $c_2$, and $c_3$ refer to their respective coefficients in the multiple linear regression.  We performed this transformation prior to all subsequent analyses, and the scaling exponents after this transformation are shown in Figure \ref{s5}. \\

\subsection{Linear Interpolation}
\begin{landscape}
	\begin{figure}[ht!]
		\centering
		\includegraphics[width=1.0\columnwidth]{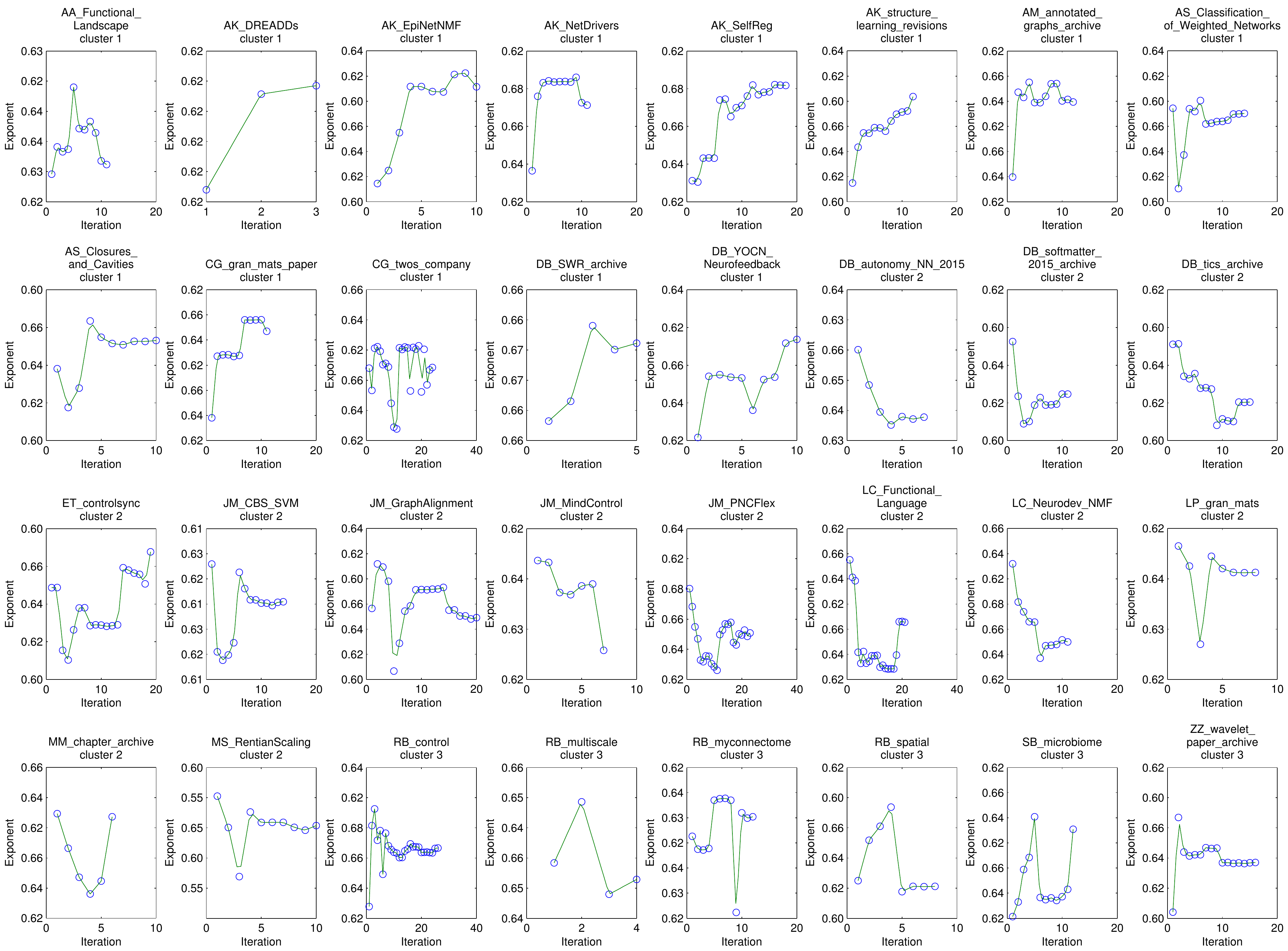}
		\caption{\textbf{Normalization of revision iterations.} Rentian scaling trends for all manuscripts (shown in blue), after correcting for the variables sparsity, density and length, along with the interpolated timeseries for each manuscript (shown in green).
			\label{s6}}
	\end{figure}
\end{landscape}

To cluster similar trends among the manuscripts using Pearson's correlation coefficient, we first interpolated the revision iterations of each manuscript using a linear interpolation. Thus, while each manuscript originally consisted of between 3 and 26 iterations of revision, after linear interpolation, each temporal trend contained the same number of data points. This enabled us to compute the Pearson's correlation coefficient between the scaling exponents over time for each pair of manuscripts. The interpolated trends are shown in Figure \ref{s6}. \\

\subsection{Lemmatization}
\begin{figure}[ht!]
	\centering
	\includegraphics[width=0.5\columnwidth]{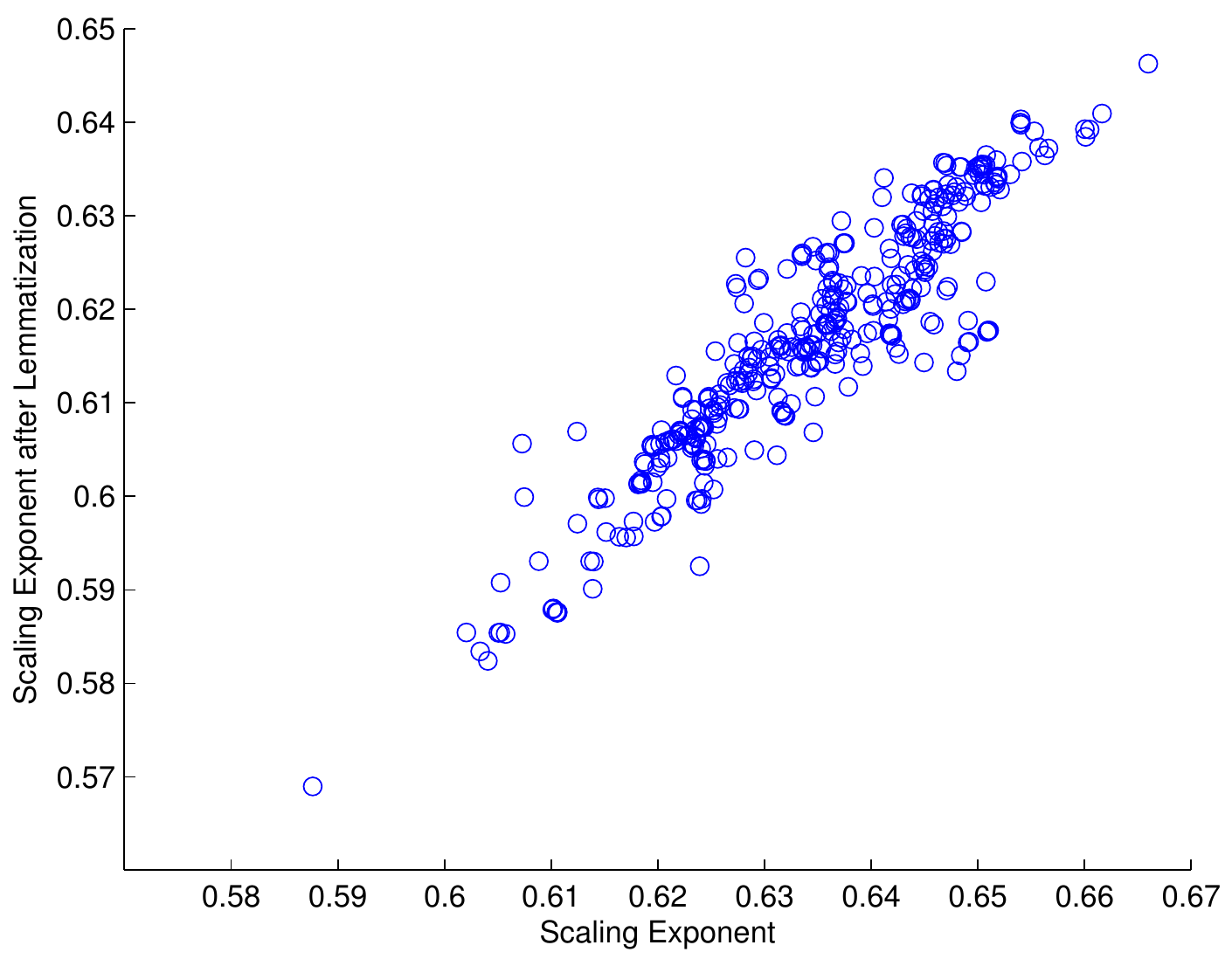}
	\caption{\textbf{Lemmatization.} We observed a high correlation between the scaling exponent without lemmatization, and the scaling exponent with lemmatization (Pearson's correlation coefficient $r=0.92, p<0.001$).
		\label{s7}}
\end{figure}

English words have many morphological variants (e.g., verb tense or plurality). In the main text, we computed the semantic networks without adjusting for morphological variations in words. However, to determine whether these morphological variants impacted network structure, we constructed the semantic networks after adjusting for these variants via lemmatization, converting all words to their base, dictionary form. We lemmatized the words by querying WordNet \cite{miller1995wordnet}, a large lexical database of English words, using the Python Natural Language ToolKit. The scaling exponent after lemmatization was highly correlated to the scaling exponent without lemmatization (Pearson's correlation coefficient $r=0.925, p<0.001$, Figure \ref{s7}), suggesting that the structure of the network remains largely the same with and without lemmatization. We further observed that lemmatization reduces the vocabulary size of the semantic networks on average by $0.054\%$, with a standard deviation of $0.011\%$.

\subsection{Changing Distance Threshold}

\begin{figure}[ht!]
	\centering
	\includegraphics[width=0.45\columnwidth]{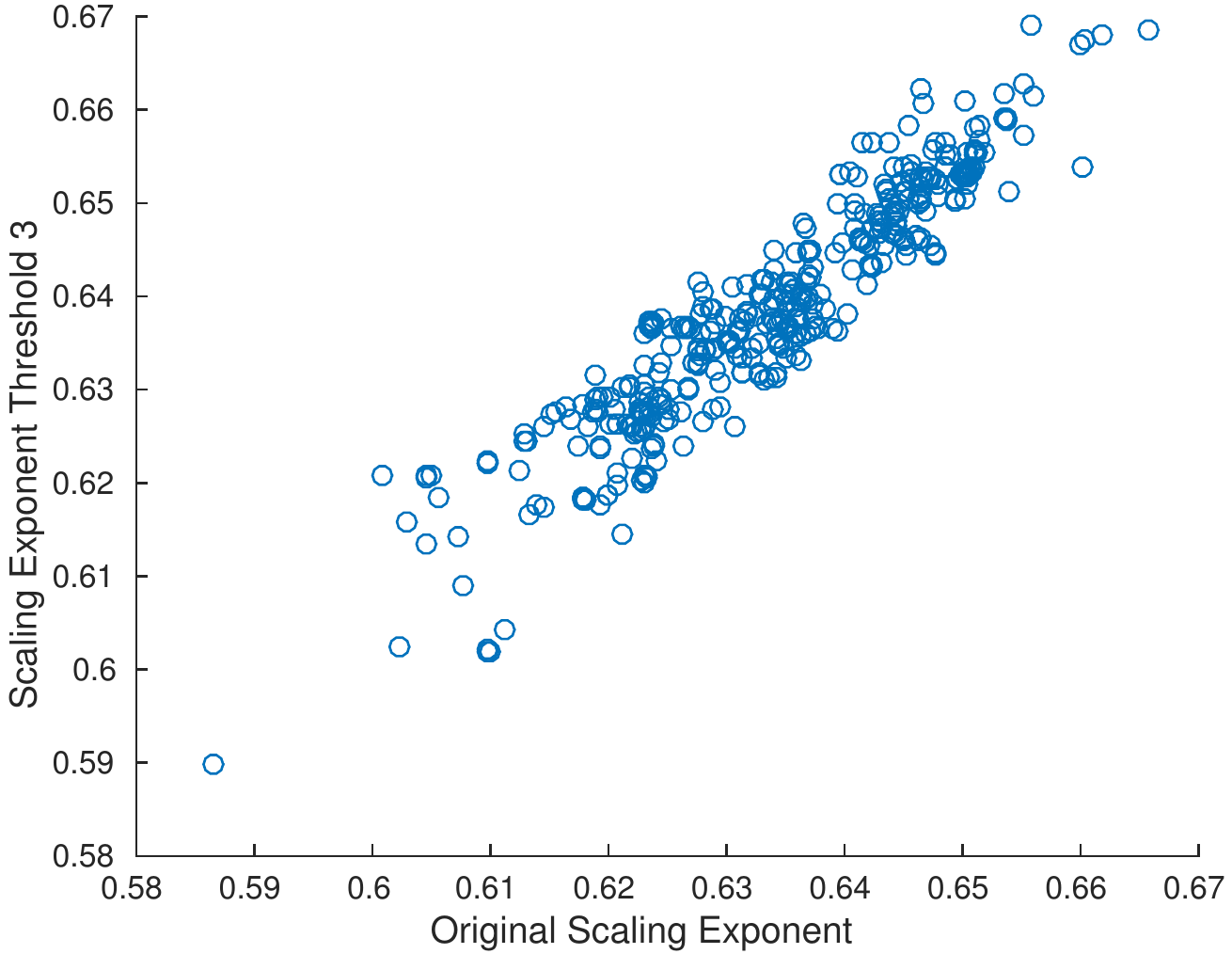}
	\includegraphics[width=0.45\columnwidth]{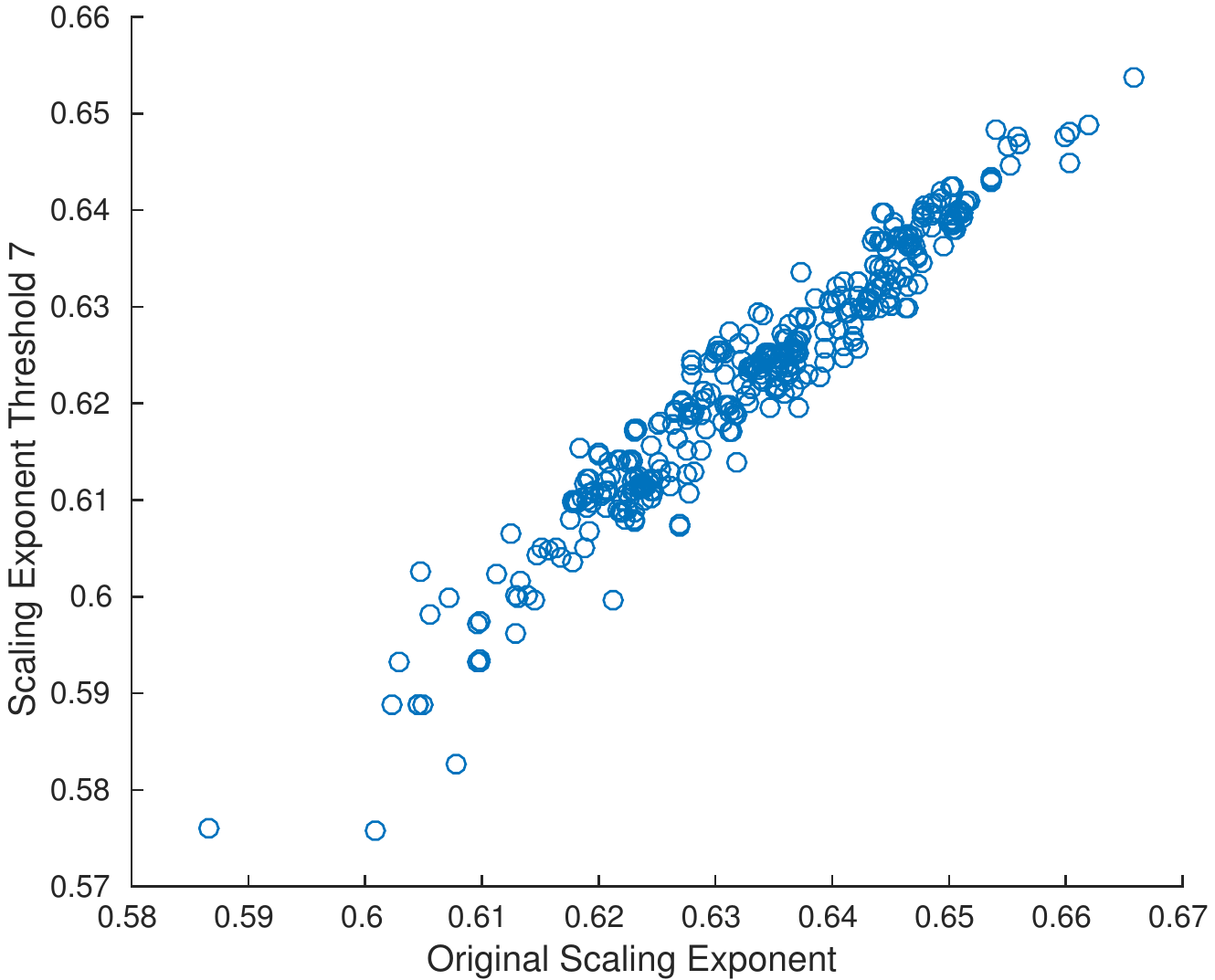}
	\caption{\textbf{Thresholding.} We observed a high correlation between the scaling exponent using a distance threshold of five words and a distance threshold of three words (left) and seven words (right).
		\label{s8}}
\end{figure}

In the main text, the distance matrices were thresholded such that pairs of words appearing within five words of one another would retain a binary edge, and words further apart would be disconnected. For robustness, scaling exponents were also computed when the threshold was three words and seven words. Comparing the three-word to the five-word cutoff, the Pearson's correlation coefficient of the scaling exponents is $r = 0.935, p < 0.001$. Comparing the seven-word to the five-word cutoff, the Pearson's correlation coefficient of the scaling exponents is $r = 0.971, p < 0.001$. See Fig. \ref{s8}. These results indicate that our findings are largely robust to small variations in this methodological choice.

\subsection{Distribution of manuscript iterations}

Manuscripts underwent between 3 and 26 iterations of revision, with a mean of 12.53, and a standard deviation of 5.81. A histogram of iteration counts is shown below in Fig.~\ref{s9}.

\begin{figure}[ht!]
	\centering
	\includegraphics[width=0.5\columnwidth]{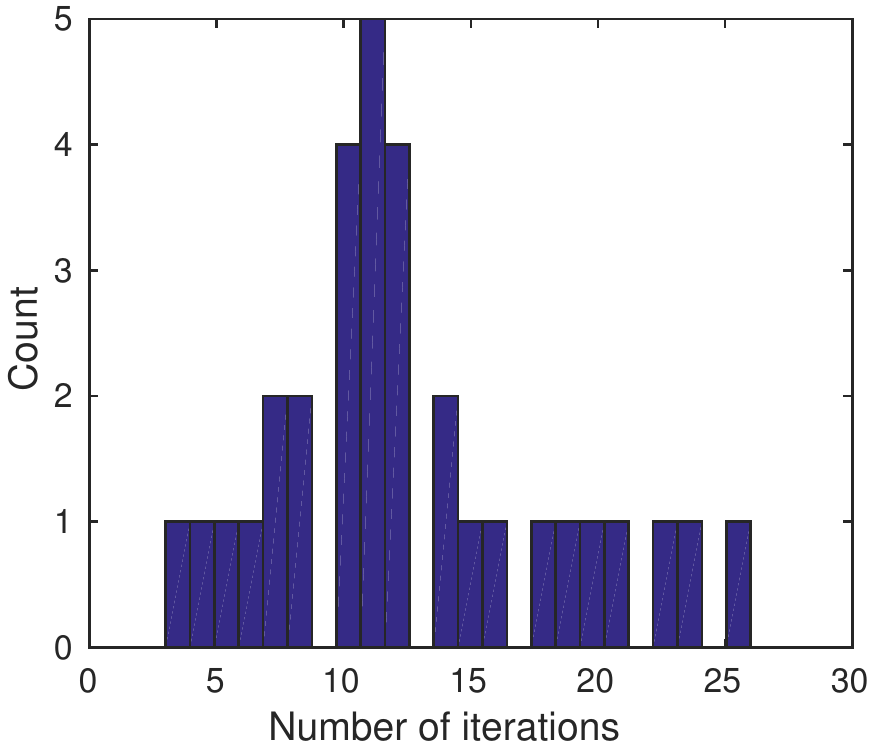}
	\caption{\textbf{Distribution of number of iterations.} Manuscripts underwent between 3 and 26 iterations of revision.
		\label{s9}}
\end{figure}

\clearpage
\newpage
\bibliographystyle{ieeetr}
\bibliography{./bibfile}